# Detecting community structure in networks using edge prediction methods

Bowen Yan[1(a)] and Steve Gregory[1]

[1] *Department of Computer Science, University of Bristol – Bristol BS8 1UB, England*

**Abstract** – Community detection and edge prediction are both forms of *link mining*: they are concerned with discovering the relations between vertices in networks. Some of the vertex similarity measures used in edge prediction are closely related to the concept of community structure. We use this insight to propose a novel method for improving existing community detection algorithms by using a simple vertex similarity measure. We show that this new strategy can be more effective in detecting communities than the basic community detection algorithms.

**Introduction.** – Many complex systems can be represented as networks, with vertices for individuals and edges describing relations between them. Analysing networks allows us to understand the structure and properties of them. In particular, community detection and edge prediction, which are both classed as link mining, according to Ref. [1], have attracted much attention in recent research.

Communities are groups of vertices in a network, such that edges between vertices in the same community are dense, but are sparse between different communities [2]. A variety of community detection algorithms [3] have been invented in recent years, involving different types of networks and community structures. For instance, the Kernighan-Lin algorithm [4] is a heuristic algorithm for partitioning networks; the Girvan-Newman algorithm [2] uses edge betweenness to select edges that should be removed between communities; modularity optimization algorithms [5-8] use a function that measures the quality of an entire partition, and tries to optimize it; label propagation algorithms [9,10], and so forth. Most algorithms are intended to detect communities in unweighted, undirected, unipartite networks, because experimental networks may not have much information for vertices and edges themselves, such as attributes and weights.

Edge prediction is the problem of predicting an edge that should exist in a network or may exist in the future. This often considers the similarity of vertices. Techniques for vertex similarity [11] from graph theory often focus on the common features that pairs of vertices share: for instance, vertex neighbourhoods, based on the local structure of networks, or all paths between vertices, based on the global structure of networks. These measures can assign a score to each pair of vertices; the higher the score, the more likely two vertices are to be neighbours.

In addition, we have found that it is effective and efficient to predict missing edges by exploiting the concept of community structure [12]. In this paper, we explore the hypothesis that it is possible to enhance the community structure of a network with extra information, *i.e.*, weights, computed by the vertex similarity measures, even if we only know the structure of the network.

In this paper, we propose a novel method that improves the existing community detection algorithms by using a vertex similarity measure based on the local structure of a network. In the next section, we present the description of our new method. Then we test this method on some artificial networks and real-world networks. Finally, we give our conclusions. In the paper, we restrict our attention to unweighted, undirected, and unipartite networks.

**Detecting communities using an edge prediction method.** – Edge prediction methods are used not only to predict edges, but to discover edges that exist in the network as well. Therefore, the first phase of our method is to use an edge prediction method to reveal the strength of the relations between vertices in order to obtain extra information in the form of weights. Then, existing community detection algorithms are used on the weighted networks.

For the first step of our method, we have chosen a simple vertex similarity measure (Common Neighbours) to calculate the scores for the existing edges in a network (Fig. 1(a)). For each edge {$u,v$}, CN($u,v$) is the number of neighbours that $u$ and $v$ have in common. Because CN($u,v$) might be zero, we add 1 to obtain the weight for edge {$u,v$} (Fig. 1(c)). The advantages of this vertex similarity measure are: (1) it focuses on the local structure of the network, which is suitable for the property of community structure; (2) compared with other vertex similarity measures, it has good performance. For the second phase, we have used some existing community detection algorithms: RFT [13], CNM [5], Infomap [14], COPRA [9], and the Louvain method [6]. These algorithms are efficient and can deal with both unweighted and weighted networks, which is important because the modified networks are weighted.

[a]E-mail: yan@cs.bris.ac.uk

**Experimental methodology.** – To evaluate a community detection algorithm, it should be tested on artificial and real-world networks. For artificial networks, we use the normalized mutual information (NMI) measure [15] to compare the known partition with the partition found by each algorithm. For real-world networks, since we do not know the real community structure, we use the modularity measure [16] to assess the quality of a partition.

*Edge prediction method and community detection algorithms used.* In the first step, a vertex similarity measure has been used to weight the original network. Here we define $score(u, v)$ to be the value of the relationship between $u$ and $v$, which is calculated in this way.

Common neighbours (CN). The number of common neighbours that two vertices have suggests the strength of the relationship between them. For example, it may be more likely that two people know each other if they have one or more acquaintances in common in a social network [17]. The function is defined as [11]

$$score(u,v) = |\Gamma(u) \cap \Gamma(v)| \qquad (1)$$

where $\Gamma(u)$ and $\Gamma(v)$ represent the set of neighbours of vertex $u$ and $v$, respectively.

In the second step, we have chosen five community detection algorithms based on different principles of graph partition, all of which can work with weighted networks. All of them except CNM automatically detect the "correct" number of communities.

The fine-tuning by reposition (or RFT) algorithm [13]. It is a fast fine-tuning algorithm, which is similar to a Kernighan-Lin optimization.

The fast greedy modularity optimization (or CNM) algorithm [5]. This algorithm begins with a trivial partition, with very low modularity, in which each vertex is a separate community. It then merges the pair of communities that results in the greatest increase in modularity.

The fast unfolding (or Louvain) method [6] is another algorithm to optimize modularity.

Infomap [14]. This algorithm uses the random walks for information flows, and detects community structure by compressing them.

COPRA [9]. This is based on the label propagation concept of Raghavan [10]. The algorithm initializes every vertex with a unique label, which then propagate between neighbours. After several iterations, the vertices in the same community have the same label.

*Network datasets used.* First, we use the LFR benchmark networks of Lancichinetti *et al.* [18]. These are artificial networks that are claimed to reflect the important aspects of real-world networks. The networks have several parameters. $n$ is the number of vertices; $\langle k \rangle$ and $k_{max}$ are the average and maximum degree; $\tau_1$ and $\tau_2$ are the exponents of the power-law distribution of vertex degrees and community sizes; $c_{min}$ and $c_{max}$ are the minimum and maximum community size. $\mu$ is the mixing parameter: each vertex shares a fraction $\mu$ of its edges with vertices in other communities.

Second, we tested some real-world networks, listed in Table 1.

Table 1. Real-world networks.

| Networks | Ref. | Type | Vertices | Edges |
|---|---|---|---|---|
| netscience | [19] | collaboration | 379 | 914 |
| email | [20] | social | 1133 | 5451 |
| scientometrics | [21] | citation | 2678 | 10368 |
| blogs | [22] | social | 3982 | 6803 |
| erdös1997 | [23] | collaboration | 5482 | 8972 |
| erdös1998 | [23] | collaboration | 5816 | 9505 |
| erdös1999 | [23] | collaboration | 6094 | 9939 |
| erdös2002 | [21] | collaboration | 6927 | 11850 |
| PGP | [24] | social | 10680 | 24316 |
| cond-mat-2003 | [25] | collaboration | 27519 | 116181 |

*Experiments with artificial networks.* We compare our method with the basic community detection algorithms that are also used in the second phase of our method. All results are averaged over 100 artificial networks with the same set of parameters. Four sets of parameters have been tested, with $\mu$ ranging from 0 to 1:

1. $n$=1000, $\langle k \rangle$=6, $k_{max}$=15, $\tau_1$=2, $\tau_2$=1, $c_{min}$=5, $c_{max}$=10.
2. $n$=1000, $\langle k \rangle$=6, $k_{max}$=15, $\tau_1$=2, $\tau_2$=1, $c_{min}$=10, $c_{max}$=20.
3. $n$=1000, $\langle k \rangle$=10, $k_{max}$=25, $\tau_1$=2, $\tau_2$=1, $c_{min}$=5, $c_{max}$=10.
4. $n$=1000, $\langle k \rangle$=10, $k_{max}$=25, $\tau_1$=2, $\tau_2$=1, $c_{min}$=10, $c_{max}$=20.

Figure 2 shows comparisons of our method with the basic community detection algorithms on LFR networks. We use four pairs of curves to represent the results for the four different sets of parameters, and each plot shows one community detection algorithm. In the figure, our method (the hollow shapes) performs better than the basic community detection algorithms (the solid shapes), except that the basic Infomap is occasionally better than our method.

*Experiments with real-world networks.* We examine the modularity obtained by the same algorithms on the real-world networks listed in Table 1. CNM does not attempt to detect the "correct" number of communities, and we do not know the real number of communities in real-world networks, so we choose the number that maximizes the modularity when the basic CNM algorithm is used, and also use the same number of communities for our method (CNM+CN). The results in Table 2 show that our methods generally compute a partition

Table 2. Real-world networks: Modularity.

| Networks | RFT | RFT+CN | CNM | CNM+CN | Infomap | Infomap+CN | COPRA | COPRA+CN | Louvain | Louvain+CN |
|---|---|---|---|---|---|---|---|---|---|---|
| netscience | 0.720 | 0.759 | 0.837 | 0.844 | 0.817 | 0.822 | 0.752 | 0.771 | 0.721 | 0.755 |
| email | 0.490 | 0.517 | 0.512 | 0.546 | 0.534 | 0.522 | 0.237 | 0.489 | 0.490 | 0.500 |
| scientometrics | 0.514 | 0.528 | 0.540 | 0.597 | 0.549 | 0.540 | 0.376 | 0.535 | 0.597 | 0.597 |
| blogs | 0.605 | 0.732 | 0.850 | 0.856 | 0.798 | 0.803 | 0.447 | 0.573 | 0.594 | 0.716 |
| erdös1997 | 0.569 | 0.696 | 0.699 | 0.726 | 0.650 | 0.703 | 0.063 | 0.388 | 0.570 | 0.682 |
| erdös1998 | 0.566 | 0.701 | 0.706 | 0.730 | 0.653 | 0.704 | 0.050 | 0.387 | 0.575 | 0.686 |
| erdös1999 | 0.568 | 0.705 | 0.699 | 0.732 | 0.655 | 0.706 | 0.050 | 0.393 | 0.571 | 0.677 |
| erdös2002 | 0.544 | 0.676 | 0.673 | 0.691 | 0.627 | 0.679 | 0.081 | 0.142 | 0.566 | 0.673 |
| PGP | 0.730 | 0.781 | 0.855 | 0.870 | 0.814 | 0.814 | 0.618 | 0.681 | 0.856 | 0.865 |
| cond-mat-2003 | 0.612 | 0.653 | 0.676 | 0.738 | 0.673 | 0.677 | 0.618 | 0.638 | 0.723 | 0.736 |

with a higher modularity than the basic community detection algorithms.

**Conclusions.** – We have proposed a simple method to detect communities in a network, which introduces the technique of edge prediction into community detection. This is because some measures of vertex similarity for edge prediction are nicely relevant to the definition of community structure. With the lack of network information, we expect to find out more useful information for edges before detecting communities. We use a simple vertex similarity measure to add weights to an unweighted network first, and then detect communities on the weighted network by using existing community detection algorithms.

We have tested our method using several efficient and effective community detection algorithms. The results on both artificial networks and real-world networks show that our strategy almost always detects communities more accurately. In addition, this vertex similarity measure relies on the local structure of a network, which is very fast, so our method does not reduce the speed of the community detection algorithms.

Our method not only supports detection of communities, but also exploits a novel idea in link mining. Many community detection algorithms consider the topology of a network while neglecting other information about vertices and edges. Indeed, this information might be missing or omitted during the process of collecting the network data. Our experiments suggest that we can use edge prediction techniques to find them. Not only that, it should be possible in future to extend our strategy to different types of networks, for example, weighted or bipartite networks, by using relevant edge prediction methods.

REFERENCES

[1] GETOOR L. and DIEHL C. P., in *Proceedings of the ACM SIGKDD International Conference on Knowledge Discovery and Data Mining* (ACM Press, New York) 2005, pp. 3-12.
[2] GIRVAN M. and NEWMAN M. E. J., *Proc. Natl. Acad. Sci. USA*, **99** (2002) 7821.
[3] FORTUNATO S., *Phys. Rep.*, **486** (2010) 75.
[4] KERNIGHAN B. W. and LIN S., *Bell. Syst. Tech. J.*, **49** (1970) 291.
[5] CLAUSET A., NEWMAN M. E. J. and MOORE C., *Phys. Rev. E*, **70** (2004) 066111.
[6] BLONDEL V. D., GUILLAUME J-L., LAMBIOTTE R. and LEFEBVRE E., *J. Stat. Mech.*, **P10008** (2008).
[7] WAKITA K. and TSURUMI T., in *Proceedings of the IADIS international conference on WWW/Internet 2007* (IADIS Press) 2007, pp. 153-162.
[8] YAN B. and GREGORY S., in *Proceedings of IEEE International Conference on Intelligent Computing and Intelligent Systems* (IEEE Press) 2009, pp. 832-836.
[9] GREGORY S., *New J. Phys.*, **12** (2010) 103018.
[10] RAGHAVAN U. N., ALBERT R. and KUMARA S., *Phys. Rev. E*, **76** (2007) 036106.
[11] LIBEN-NOWELL D. and KLEINBERG J., *J. Am. Soc. Inf. Sci. Tec.*, **58** (2007) 1019.
[12] YAN B. and GREGORY S., arXiv: 1109.2793, 2011.
[13] GRANELL C., GÓMEZ S. and ARENAS A., *Chaos*, **21** (2011) 016102.
[14] ROSVALL M. and BERGSTROM C. T., *Proc. Natl. Acad. Sci. USA*, **105** (2008) 1118.
[15] LANCICHINETTI A., FORTUNATO S. and KERTÉSZ J., *New J. Phys.*, **11** (2009) 033015.
[16] NEWMAN M. E. J. and GIRVAN M., *Phys. Rev. E*, **69** (2004) 026113.
[17] NEWMAN M. E. J., *Phys. Rev. E*, **64** (2001) 025102.
[18] LANCICHINETTI A., FORTUNATO S. and RADICCHI F., *Phys. Rev. E*, **78** (2008) 046110.
[19] NEWMAN M. E. J., *Phys. Rev. E*, **74** (2006) 036104.
[20] GUIMERA R., DANON L., DIAZ-GUILERA A., GIRALT F. and ARENAS A., *Phys. Rev. E*, **68** (2003) 065103.
[21] PAJEK, Network from Pajek datasets. http://vlado.fmf.uni-lj.si/pub/networks/data/
[22] XIE N., *Social network analysis of blogs*, unpublished, 2006.
[23] BATAGELJ V. and MRVAR A., *Soc. Networks*, **22** (2000) 173.
[24] BOGUŇÁ M., PASTOR-SATORRAS R., DÍAZ-GUILERA A. and ARENAS A., *Phys. Rev. E*, **70** (2004) 056122.
[25] NEWMAN M. E. J., *Proc. Natl. Acad. Sci. USA*, **98** (2001) 404.

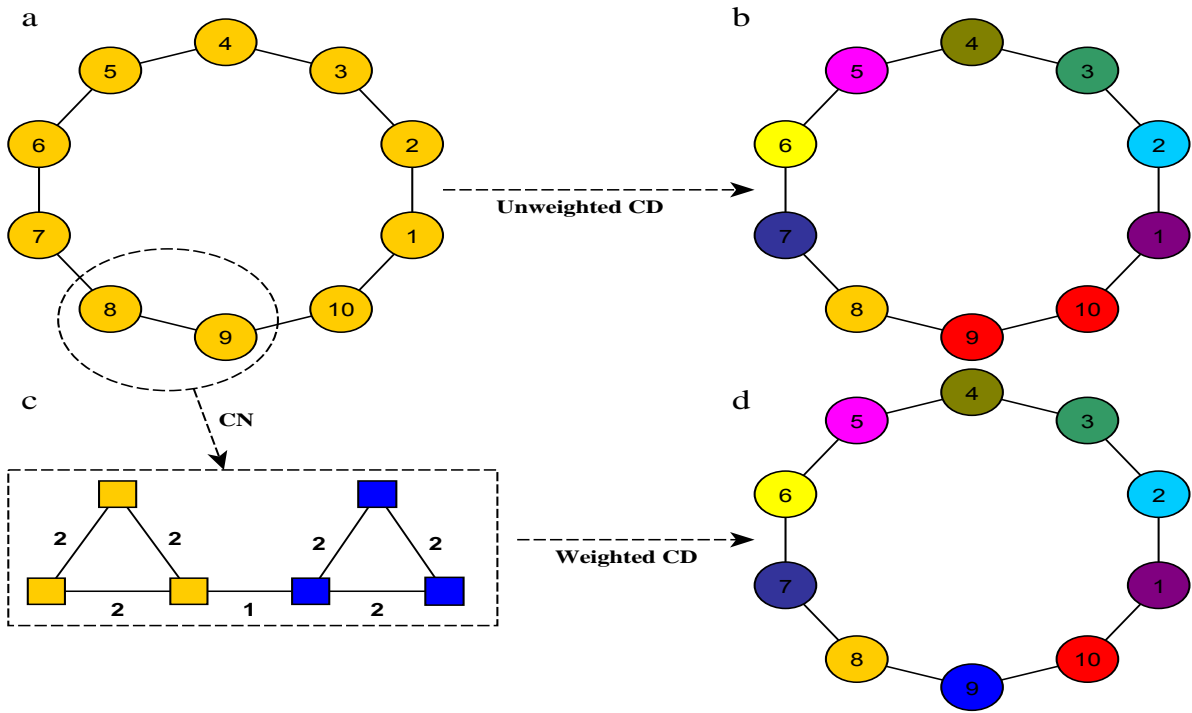

Fig. 1: The mechanism of the method. a) The original network. Each circle represents a clique, and each pair of cliques is connected by one edge. b) Communities obtained by Infomap from the original network, 9 and 10 should be two communities rather than one community. c) Edges are weighted by the CN method. d) Correct communities obtained by Infomap from the weighted network.

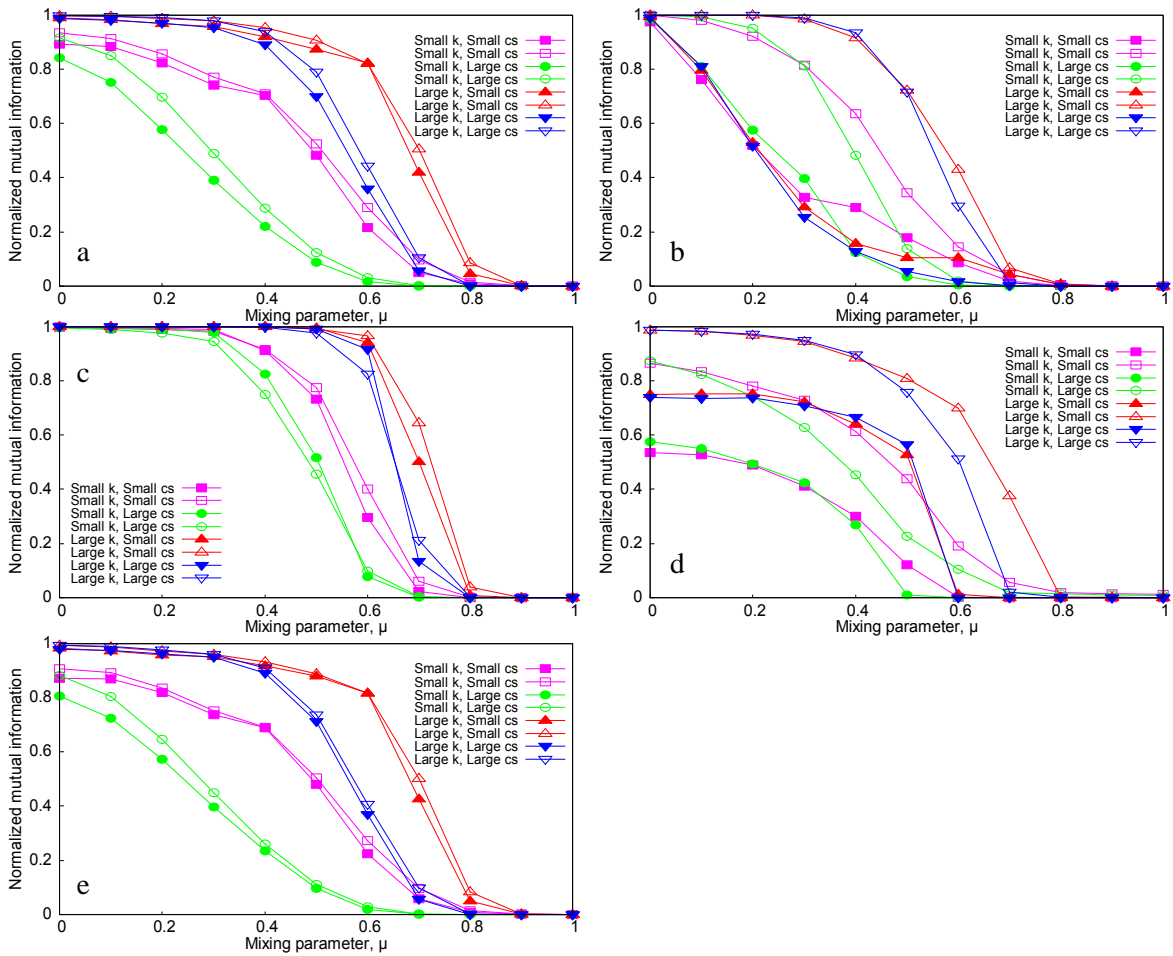

Fig. 2: Performance of our method and the basic community detection algorithms on LFR networks. a) RFT. b) CNM. c) Infomap. d) COPRA. e) Louvain method.